\documentclass[aps,prl,reprint,a4paper,floatfix,amsmath,amssymb,amsfonts,noshowpacs]{revtex4-2}
\usepackage{newtxtext,newtxmath}
\usepackage[utf8]{inputenx}
\usepackage{graphicx}
\usepackage[dvipsnames]{xcolor}
\usepackage{chemist}
\usepackage[pdftex,unicode=true,bookmarks=false,breaklinks=false,pdfborder={0 0 1},colorlinks=true]{hyperref}
\hypersetup{linkcolor=blue,citecolor=blue,urlcolor=blue}
\usepackage[textwidth=17.5cm,textheight=23.5cm,verbose,pdftex]{geometry}
\usepackage{soul}
\usepackage{tikz}
\usepackage{chemfig}
\usepackage{mhchem}
\usetikzlibrary{calc,fadings,decorations.markings}

\newcommand{\fga}{$\phi_{\text{Ga}}$}
\newcommand{\fo}{$\phi_{\text{O}}$}

\newcommand{\bgao}{$\upbeta$-\text{Ga}$_2$\text{O}$_3$}
\newcommand{\balgao}{$\upbeta$-\text{(Al,Ga)}$_2$\text{O}$_3$}
\newcommand{\aalgao}{$\upalpha$-\text{(Al,Ga)}$_2$\text{O}$_3$}
\newcommand{\agao}{$\upalpha$-\text{Ga}$_2$\text{O}$_3$}
\newcommand{\aalo}{$\upalpha$-\text{Al}$_2$\text{O}$_3$}

\newcommand{\tg}{$T_{\text{G}}$}
\newcommand{\gr}{$\varGamma$}

\newcommand{\pga}{$\,\text{nm}^{-2}\,\text{s}^{-1}$}
\newcommand{\pin}{$\,\text{nm}^{-2}\,\text{s}^{-1}$}

\newcommand{\pfu}{$\text{nm}^{-2}\,\text{s}^{-1}$}

\newcommand{\utg}{$\,^{\circ}\text{C}$}
\newcommand{\utge}{$^{\circ}\text{C}$}
\newcommand{\gru}{$\,\text{nm}\,\text{min}^{-1}$}
\newcommand{\sccm}{$\,\text{SCCM}$}

\newcommand{\uni}{$\,\text{nm}^{-2}\,\text{s}^{-1}$}
\newcommand{\fin}{$\phi_{\text{In}}$}

\begin{document}





\graphicspath{{./figs/}}

\title{Growth of \agao\ on \aalo\ by conventional molecular-beam epitaxy \\ and metal-oxide-catalyzed epitaxy}
\author{J. P. McCandless}
\email[Electronic mail: ]{jpm432@cornell.edu}
\affiliation{School of Electrical and Computer Engineering, Cornell University, Ithaca, New York 14853, USA}
\author{D. Rowe}
\affiliation{Department of Material Science and Engineering, Cornell University, Ithaca, New York 14853, USA}
\author{N. Pieczulewski}
\affiliation{Department of Material Science and Engineering, Cornell University, Ithaca, New York 14853, USA}
\author{V. Protasenko}
\affiliation{School of Electrical and Computer Engineering, Cornell University, Ithaca, New York 14853, USA}
\author{M. Alonso-Orts}
\affiliation{Institute of Solid-State Physics, University Bremen, Otto-Hahn-Allee 1, 28359 Bremen, Germany}
\author{M. S. Williams}
\affiliation{Institute of Solid-State Physics, University Bremen, Otto-Hahn-Allee 1, 28359 Bremen, Germany}
\author{M. Eickhoff}
\affiliation{Institute of Solid-State Physics, University Bremen, Otto-Hahn-Allee 1, 28359 Bremen, Germany}
\author{H. G. Xing}
\affiliation{School of Electrical and Computer Engineering, Cornell University, Ithaca, New York 14853, USA}
\affiliation{Department of Material Science and Engineering, Cornell University, Ithaca, New York 14853, USA}
\affiliation{Kavli Institute at Cornell for Nanoscale Science, Cornell University, Ithaca, New York 14853, USA}
\author{D. A. Muller}
\affiliation{Kavli Institute at Cornell for Nanoscale Science, Cornell University, Ithaca, New York 14853, USA}
\affiliation{School of Applied and Engineering Physics, Cornell University, Ithaca, New York 14853, USA}
\author{D. Jena}
\affiliation{Department of Material Science and Engineering, Cornell University, Ithaca, New York 14853, USA}
\affiliation{School of Electrical and Computer Engineering, Cornell University, Ithaca, New York 14853, USA}
\affiliation{Kavli Institute at Cornell for Nanoscale Science, Cornell University, Ithaca, New York 14853, USA}
\author{P. Vogt}
\affiliation{Department of Material Science and Engineering, Cornell University, Ithaca, New York 14853, USA}
\affiliation{Institute of Solid-State Physics, University Bremen, Otto-Hahn-Allee 1, 28359 Bremen, Germany}

\begin{abstract}
\noindent
We report the growth of \agao\ on $m$-plane \aalo\ by conventional plasma-assisted molecular-beam epitaxy (MBE) and In-mediated metal-oxide-catalyzed epitaxy (MOCATAXY). We report a growth-rate-diagram for \agao($10\Bar{1}0$), and observe (i) a growth rate increase, (ii) an expanded growth window, and (iii) reduced out-of-lane mosaic spread when MOCATAXY is employed for the growth of \agao. Through the use of In-mediated catalysis, growth rates over $0.2\,\upmu\text{m}\,\text{hr}^{-1}$ and rocking curves with full width at half maxima of $\Delta\omega \approx 0.45^{\circ}$ are achieved. Faceting is observed along the \agao film surface and is explored through scanning transmission electron microscopy. 
\end{abstract}

\maketitle
\section{Introduction}
\noindent
Over the past decade, \ce{Ga2O3} has gained much attention as a wide-band gap semiconductor. Monoclinic \bgao\ possesses an ultra-wide bang gap of $\sim 4.7\,\text{eV}$ \cite{Tippins1965}, and it has been the most studied phase owing to its thermal stability and the availability of large-area, native, semi-insulating and conductive substrates \cite{Galazka2017, Kuramata2016_substrates}. To further increase its band gap \bgao\ can be alloyed with Al to form \balgao, but achieving high Al content has remained challenging due to the tendency to have phase segregation \cite{Bhuiyan2020}. In contrast, \aalgao\ becomes more stable as the Al is increased because the crystal is isostructural with the \aalo\ substrate, and the lattice mismatch is reduced as the Al concentration is increased \cite{Jinno2021}. This has enabled the entire compositional range of \aalgao\ to be readily achieved \cite{Jinno2021,Bhuiyan2021_alphGO}, and it has enabled band gaps exceeding those of AlN, BN, or diamond \cite{Tsai2009,Cassabois2016}.  

\noindent
With the recent advances enabling \agao\ to remain stable during high-temperature anneals \cite{mccandless2021}, the next challenge is to achieve electrical conductivity. To date, conductivity in \agao\ has been achieved by chemical vapor deposition (CVD) \cite{Akaiwa2012,uchida2018}, but has remained elusive for films grown by molecular-beam epitaxy (MBE). Additionally, conductive \bgao\ films grown by MBE on \aalo\ have yet to be achieved \cite{Tsai2009}. While the exact reasons these films remain insulating are unknown, the thermodynamics during MBE growth and the low formation energy of defects may cause these \ce{Ga2O3} films on \ce{Al2O3} to be insulating. 

\noindent
Multiple compensating point defects (e.g., cation vacancies, oxygen vacancies, donor impurities \cite{Varley2010,Varley2011}) and extended crystallographic defects (e.g., rotational domains and threading dislocations\cite{Rafique2018}) occur within the \ce{Ga2O3} films grown on sapphire. Reasons for the emergence of these defects include the lattice mismatch between the film and the substrate \cite{Rafique2018}, and the growth regime in which the films are grown \cite{Varley2010,Varley2011,Korhonen2015}. For example, Ga vacancies ($V_{\text{Ga}}$) may act as compensating acceptors for introduced $n$-type dopants in grown \ce{Ga2O3} thin films \cite{Korhonen2015}. O-rich growth environments are likely to generate a significant amount of $V_{\text{Ga}}$ (due to their low formation energy) whereas Ga-rich growth environments are found to significantly  suppress the formation of $V_{\text{Ga}}$ (due to their high formation energy) \cite{Varley2011}. Thus, the growth of \ce{Ga2O3} in the highly Ga-rich regime---accessed by new epitaxial growth concepts \cite{Vogt2022_PhysRevApp}---may improve the transport properties of \ce{Ga2O3} thin films since the Ga-rich growth regimes lead to higher $V_{\text{Ga}}$ formation energies, resulting in  lower $V_{\text{Ga}}$ densities within the \ce{Ga2O3} layers.

\noindent
One approach to address these issues is through the use of metal-oxide-catalyzed epitaxy (MOCATAXY) \cite{Vogt2018_MOCATAXY}. This is a growth process where a catalytic element (e.g. In) is introduced to the growth system and results in metal-exchange catalysis \cite{Vogt2017_mocataxy}. This growth mode has been observed for \balgao\ on different substrates and surface orientations, as well as for different epitaxial growth techniques \cite{Mazzolini2019,Mauze2020, Mazzolini2020,Kuang2021,Kracht2017}. 

\noindent
Many benefits arise from using MOCATAXY during the growth of \ce{Ga2O3}. For example:~(i) It can improve the surface morphologies of \bgao-based films \cite{Mauze2020}. (ii) The synthesis of \ce{Ga2O3} can occur in previously inaccessible kinetic and thermodynamic growth regimes (e.g. in highly metal-rich regimes) which can be beneficial for the suppression of undesired point (such as $V_{\text{Ga}}$) defects in \ce{Ga2O3} \cite{Vogt2017_mocataxy,Varley2010,Varley2011}. (iii) The formation of thermodynamically unstable \ce{Ga2O3} phases becomes energetically favorable \cite{Kracht2017,Vogt2017_mocataxy, Vogt2022_PhysRevApp}, e.g., the formation of the $\upepsilon/\upkappa$-phase of \ce{Ga2O3}, which has enabled novel $\upepsilon/\upkappa$-\ce{Ga2O3}-based heterostructures \cite{Kuang2021}. (iv) The growth rate (\gr), possible growth temperatures (\tg), and crystalline quality of \balgao-based thin films can be vastly enhanced \cite{Vogt2018_MOCATAXY}.

\noindent
In this work, we introduce the growth of \agao\ by MOCATAXY, resulting in an expansion of the \agao\ growth window combined with an increased \gr\ and an improvement in its out-of-plane mosaic spread. It is the first demonstration of a catalytic growth process during the growth of \agao.

\section{Experimental}
\noindent
Samples were grown in a Veeco GEN930 plasma MBE system with standard Ga and In effusion cells. For all samples, the substrates were cleaned in acetone and isopropanol for 10 minutes and the \agao\ samples were grown for 60 minutes. The growth temperature ($T_{\text{G}}$) was measured by a thermocouple located within the substrate heater. The Ga flux (\fga) and In flux (\fin) were monitored by beam equivalent pressure (BEP) chamber readings. For conventional MBE and MOCATAXY, the \ce{O2} flux (\fo) was measured in standard cubic centimeters per minute (SCCM) and a radio-frequency plasma power of $250\,\text{W}$ was employed during all growths. To convert the measured values of \fga\ (BEP), \fin\ (BEP), and \fo\ (SCCM) into units of $\text{nm}^{-2}\,\text{s}^{-1}$ conversion factors are taken from Ref.~\cite{Vogt2017_dissertation}. Note, when using In-mediated catalysis, the available \fo\ for Ga to \ce{Ga2O3} oxidation is about $2.8$ times larger than for Ga oxidation in the absence of In \cite{Vogt2017_mocataxy,Vogt2022_PhysRevApp}. 

\noindent
For samples grown by conventional MBE and MOCATAXY, the impact of \fga\ and \tg\ is studied. In the case of MOCATAXY growth, the impact of \fin\ is also investigated. All growth parameters used in this work are collected in Table~\ref{tab}.

\noindent
For scanning transmission electron microscopy (STEM), samples were prepared using Thermo Fisher Helios G4 UX Focused Ion Beam with a final milling step of 5 keV to minimize damage. Carbon and Au-Pd layers were sputtered to reduce charging during sample preparation. Carbon and platinum protective layers were also deposited to minimize ion-beam damage. STEM measurements were taken with an aberration-corrected Thermo Fisher Spectra 300 CFEG operated at 300 keV.
\begin{table}[t]
\caption{Collected growth parameters used in this work, values of \fga, \fin, \fo, and \tg, for samples grown by conventional MBE and MOCATAX are listed. The conversion for \fga\ and \fin\ from BEP to \gru\ to \pfu\ are $\text{\fga} = 2.5 \times 10^{-8} \, \text{Torr} \; \hat{=} \; 1.1 \, \text{nm}\,\text{min}^{-1} \; \hat{=} \; 1 \text{\pga}$ and $\text{\fin} = 1.1 \times 10^{-7}\, \text{Torr} \; \hat{=} \;  2.6\text{\pin}$, respectively.}
\centering
\begin{tabular}{ ||c|c|c|| } 
 \toprule
\; Growth parameters \; & \; Conventional MBE \; & \;\; MOCATAXY \;\; \\ \hline 
$\text{\fga}\,{\text{(\pfu)}}$ & 0.8 -- 2.0      &  1.1 -- 5.5        \\ \hline
$\text{\fin}\,{\text{(\pfu)}}$ &      0          &  2.6 -- 2.8        \\ \hline
$\text{\fo}\,{\text{(SCCM)}}$  &     1.4         &  0.7 -- 1.0        \\ \hline
$\text{\fo}\,{\text{(\pfu)}}$  &     2.2         &  3.2 -- 4.6        \\ \hline
$\text{\tg}\,{\text{(\utge)}}$ &  640 -- 800     &  680               \\ \hline 
\end{tabular}
\label{tab}
\end{table}

\section{Results and Discussion}
\begin{figure}[ht]
\includegraphics[scale=0.825]{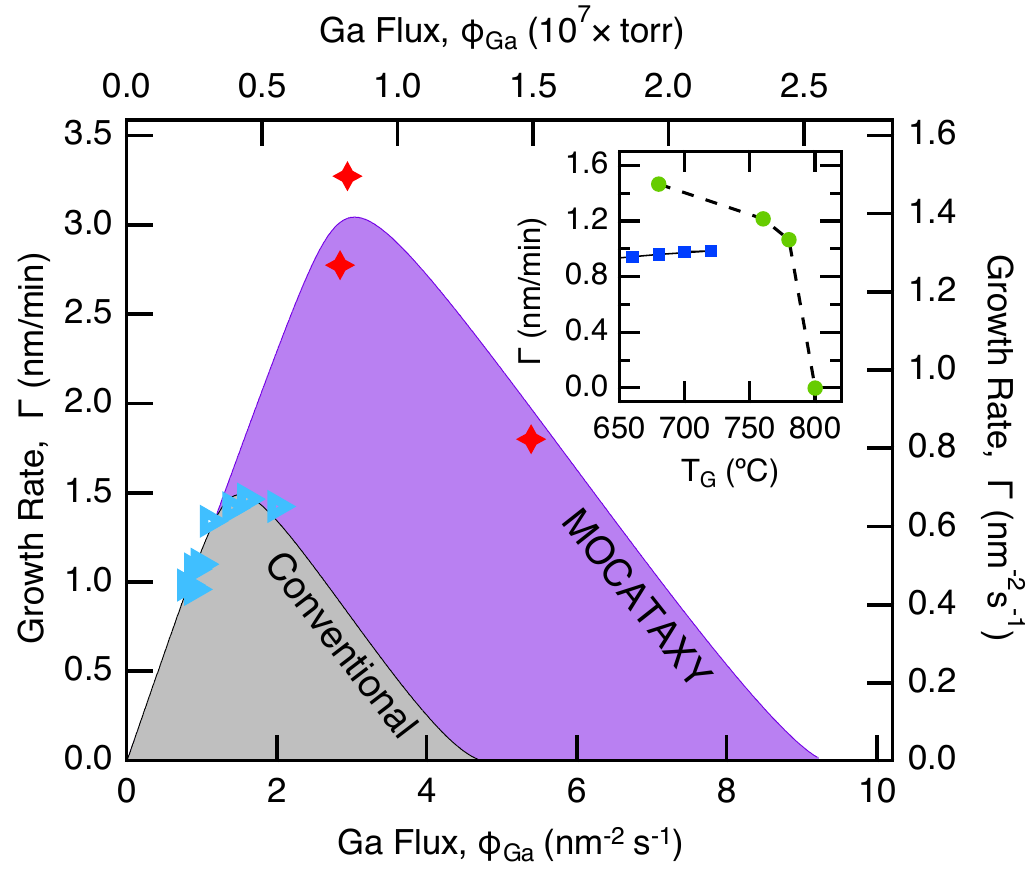}
\caption{Growth-rate-diagram of \agao($10\Bar{1}0$) grown on \aalo($10\Bar{1}0$). The growth rate \gr\ as a function of \fga\ at \tg\ = 680\utg\ is plotted for the growth by conventional MBE (blue triangles) and MOCATAXY (red stars). The \gr-data is fit by a \gr-model taken from Ref.~\cite{vogt2018_model}. The gray shaded region shows the parameter space under which the formation of \agao by conventional MBE may occur. The purple shaded area depicts the growth regime of \agao assisted by MOCATAXY. Both fitted data sets were obtained at constant \tg\ and \fo\ (values given in Table~\ref{tab}). Inset:~\gr\ as a function of \tg\ at two different fluxes of (i) \fga\ = 0.9\pga\ (the O-rich regime, solid squares) and (ii) \fga\ =  1.6\pga\ (the \gr-plateau regime, solid discs). A growth-rate-diagram of \agao\ as a function of \fo\ is given in Ref.~\cite{Note1}.}
\label{fig:growthDiagram}
\end{figure}

\noindent
Figure \ref{fig:growthDiagram} shows the growth-rate-diagram of \agao($10\Bar{1}0$) grown on \aalo($10\Bar{1}0$) by conventional MBE (the gray shaded area) and MOCATAXY (the purple shaded area). For conventionally grown samples two distinct growth regimes are observed: (i) the O-rich regime where O adsorbates are in excess over Ga adsorbates (i.e., the Ga flux limited regime), and (ii) the \gr-plateau regime (i.e., the Ga$_2$O desorption limited regime). The O-rich regime is characterized by an increasing \gr\ with increasing \fga, whereas the plateau regime is characterized by a constant \gr, being independent of \fga. Within this regime, however, \gr\ may decrease with increasing \tg\ (see inset in Fig.~\ref{fig:growthDiagram}) as the desorption of the volatile suboxide Ga$_2$O becomes thermally more active \cite{Vogt2016}. The data in the inset of Fig.~\ref{fig:growthDiagram} plot \gr\ as a function of \tg:~(i) for \agao\ grown the O-rich regime and (ii) for \agao\ grown in the \gr-plateau regime. 

\noindent
To expand the accessible growth window of \agao\ to higher \fga\ and higher \tg, combined with increased \gr\ and improved crystalline quality, In-mediated catalysis was employed to the formation of \agao\ \cite{Vogt2017_mocataxy}. The red stars in Fig.~\ref{fig:growthDiagram} show the resulting \gr\ as a function of \fga\ at constant \tg.
The gray shaded and purple shaded areas in Fig. \ref{fig:growthDiagram} depict model-based descriptions of \gr\ for \agao\ grown by conventional MBE and MOCATAXY, respectively. The maximum \gr\ obtained for each growth technique is $\text{\gr} \approx 1.5\text{\gru}$ and $\text{\gr} \approx 3.3\text{\gru}$, respectively. Using MOCATAXY, a more than 2-times increase in \gr\ for \agao\, at given growth conditions, as well as a shift far into the adsorption-controlled regime (i.e, far into the Ga rich flux regime) is observed. This direct comparison between the two growth types clearly shows the expanded growth window made possible with MOCATAXY, for example, enabling $\text{\gr} \approx 1.8\text{\gru}$ for \agao\ at \fga\ = 5.5\pga. In contrast, at these growth conditions, \textit{no} growth of \agao\ is obtained by conventional MBE. The catalytic effect on \gr\ of \agao\  is modeled as a function of \fo\ within the supplemental section \footnote{See Supplemental Material at [URL will be inserted by publisher] for a model of \gr\ as a function of \fo\ at \tg\ = 680\utg\ (S-Fig.~1) as well as images obtained by STEM (S-Fig.~2 and S-Fig.~3) and an included crystallographic model.}. We note that the depicted models use arbitrary kinetic parameters, based on kinetic parameters extracted for the growth of \bgao\ \cite{vogt2018_model}.

\noindent
To describe the growth of \agao\ by MOCATAXY, \fo\ is scaled by a factor of 2.8 compared with the growth of \agao\ by conventional MBE. This additional O comes from the catalytic nature of In forming a catalytic adlayer ($A$) with O adsorbates, e.g., $A$ = In--O, which provides more active O for the Ga to \agao\ oxidation. In other words, $A$ increases the reaction probability of Ga with O on the respective growth surface, facilitating the formation of the final \ce{Ga2O3} compound at much higher \fga\ and \tg, which enables excellent crystal quality \cite{Vogt2017_mocataxy,Vogt2022_PhysRevApp}. We further note that the same factor of 2.8 was needed for modeling the MOCATAXY growth of \bgao\ on different substrates and different surface orientations \cite{Vogt2017_mocataxy,Vogt2022_PhysRevApp}. We note, however, that for a quantitative extraction of all kinetic growth parameters more \gr-studies of \agao\ are needed and are beyond the scope of this work. Nevertheless, the models help validate the \gr-data and provide insight into the growth regimes and growth mechanisms of \agao. For example, once \fga\ exceeds the active O flux, i.e., for $\text{\fga} > \text{\fo}$, the growth will enter the Ga-rich regime and \gr\ will start to decrease, as shown by the gray shaded area in Fig.~\ref{fig:growthDiagram}. Thus, this is the first direct indication that the growth of \agao\ is limited by the formation and subsequent desorption of Ga$_2$O, like what is observed for \bgao\ grown by conventional MBE \cite{vogt2018_model}.

\begin{figure}[t]
\includegraphics[scale = 0.65]{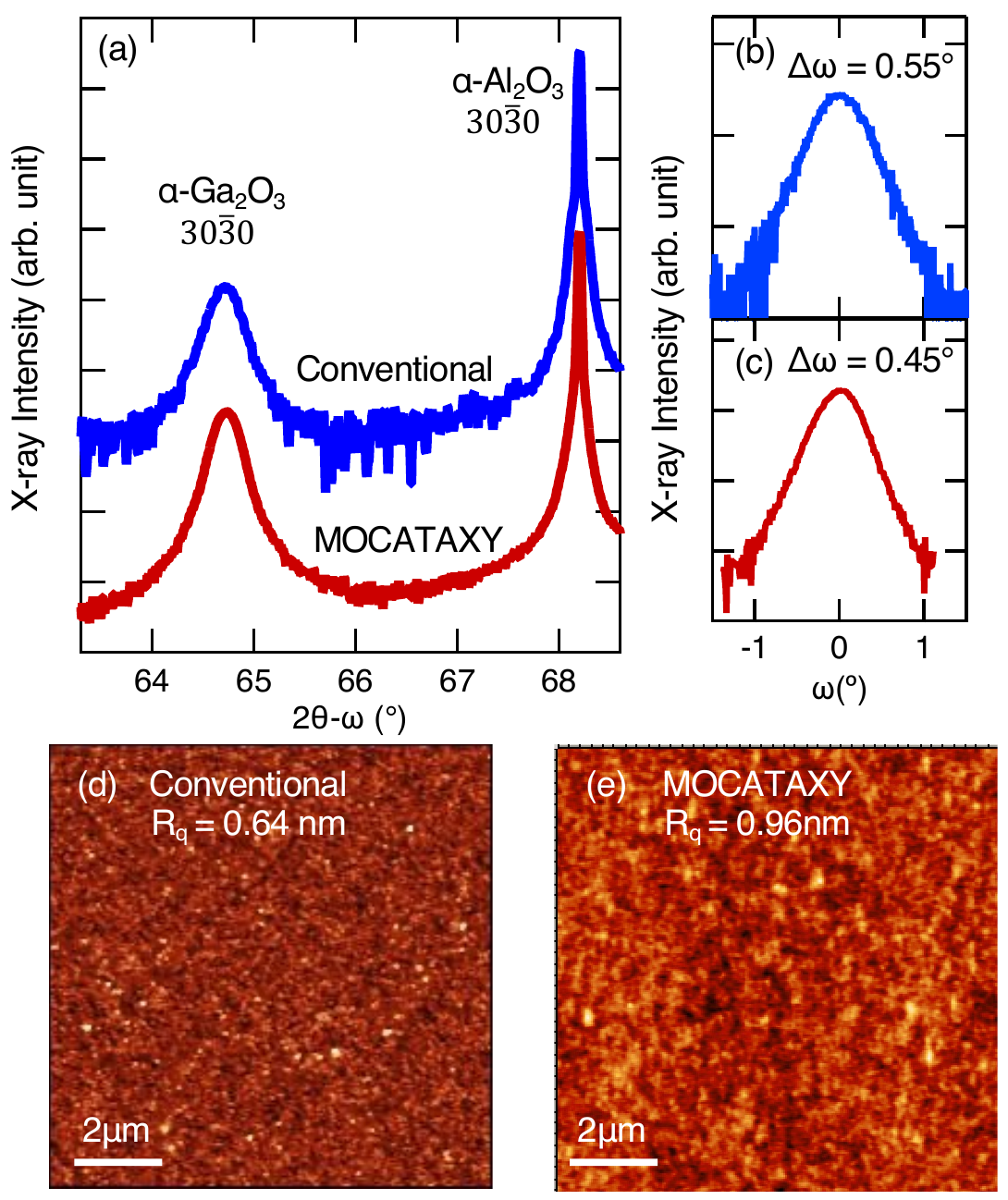}
\caption{(a) Longitudinal XRD scans of optimized  \ce{\alpha-Ga2O3} films. The reflections of the films coincide with the \ce{\alpha-Ga2O3}$(10\Bar{1}0)$ phase grown by conventional MBE (the blue trace) and MOCATAXY (the red trace). The used growth parameters were \fga\ = 2.9\uni, $\text{\fo} = 1.4\text{\sccm} \; \hat{=} \; 2.2\text{\uni}$, and \tg\ = 750\utg\ (conventional MBE), and \fga\ = 2.9\uni, \fin\ = 2.8\uni, $\text{\fo} = 0.7\text{\sccm} \; \hat{=} \; 3.2\text{\uni}$, and \tg\ = 680\utg\ (MOCATAXY). (b) and (c) Transverse XRD scans across the $30\Bar{3}0$ peak with their FWHM of $\Delta\omega = 0.55^{\circ}$ (conventionally MBE-grown) and $\Delta\omega = 0.45^{\circ}$ (MOCATAXY-grown). These obtained $\Delta \omega$ are depicted in Fig.~\ref{fig:allCompare} at given \fga\ and \tg. (d) and (e) Surface morphologies obtained by $10 \times 10\,\upmu\text{m}$ AFM scans for \agao($10\Bar{1}0$) surfaces grown by conventional MBE and MOCATAXY, respectively. Growth conditions for the samples plotted in (d) and (e) are the same as for the ones plotted in panels (a)--(c), except a slightly lower \tg = 730\utg\ used for the conventionally grown sample and a slightly higher supplied $\text{\fo} = 1.0\text{\sccm}$ for the MOCATAXY grown film. This resulted in $\Delta\omega = 0.61^{\circ}$ and $\text{\gr} \approx 1.2\text{\gru}$ for the conventionally grown sample, and $\Delta\omega = 0.48^{\circ}$ and $\text{\gr} > 3.0\text{\gru}$ for the MOCATAXY grown sample.}
\label{fig:RC_compare}
\end{figure}
\begin{figure}[t]
\includegraphics[scale = 0.675]{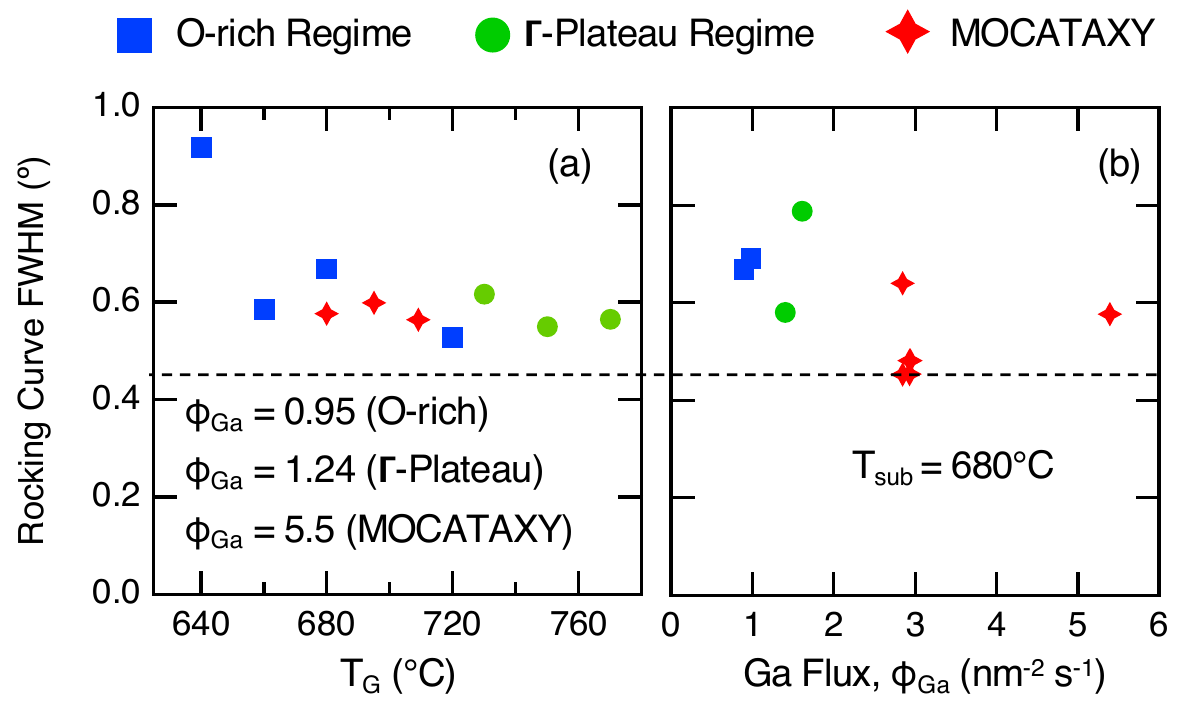}
\caption{(a) and (b) FWHM (i.e., $\Delta\omega$ values) are plotted as a function of \tg\ and \fga\ are plotted, respectively. Values are obtained by transverse XRD scans of the $30\Bar{3}0$ peaks of \agao\ grown films (XRD data not shown). Three distinct growth regimes are studied in panels (a) and (b):~(i) the O-rich rich regime (blue squares), (ii) the \gr-plateau regime (green circles), and (iii) the MOCATAXY regime (red stars). The lowest value of $\Delta\omega$ is indicated by a dashed line. Note that for the samples grown by MOCATAXY, \fin\ = (2.6 -- 2.8)\pga\ was supplied and might explain the slight variations observed in $\Delta \omega$ for \agao\ grown at \fga\ = 2.9\pga\ in panel (b)].}
\label{fig:allCompare}
\end{figure}

\begin{figure*}[t]
\includegraphics[scale = 0.575]{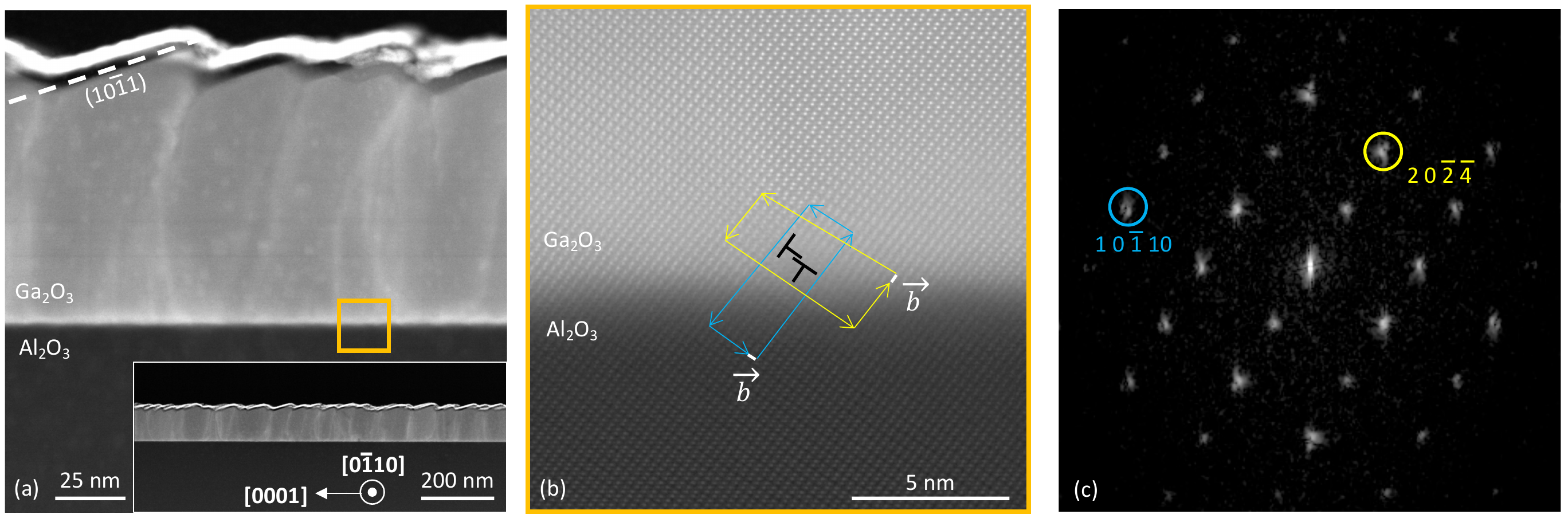}
\caption{HAADF-STEM images show an overview of an \agao($10\Bar{1}0$) film grown on \aalo($10\Bar{1}0$). (a) The epitaxial film shows increased contrast due to misfit dislocations at the film/substrate interface. Threading dislocation propagate through the film and terminating at the intersection of its surface periodic faceting. (b) Enlarged image of the film-substrate interface (i.e., the \aalo-\agao\ interface) is shown. Burger circuits are drawn around the edge dislocations. (c) Fast Fourier transform (FFT) of the interface region is shown. Diffraction peak separation at ($20\Bar{2} \Bar{4}$) and ($10\Bar{1}10$) indicate strain relaxation of the \agao($10\Bar{1}0$) on \aalo($10\Bar{1}0$).}
\label{fig:4}
\end{figure*}

\noindent
Figure \ref{fig:RC_compare} directly compares the impact of both MBE growth techniques on the structural quality of the epitaxially grown films. In Fig. \ref{fig:RC_compare} (a), $2\theta$-$\omega$ XRD scans of two selected \agao\ films are shown, one grown by conventional MBE (depicted as the blue trace) and one grown by MOCATAXY (depicted as the red trace). The reflections of the films coincide with the \agao\ $30\Bar{3}0$ peak. This, along with the absence of other diffraction peaks, indicates phase-pure \agao(10$\bar{1}$0) with In incorporation of $< 1 \%$ in the grown \agao\ layers, similar to what is observed for \balgao\ grown by MOCATAXY \cite{Vogt2018_MOCATAXY}. Fig. \ref{fig:RC_compare}(b) and \ref{fig:RC_compare}(c) plot transverse scans (rocking curves) for the conventional MBE and MOCATXY grown \agao\ samples as plotted in Fig.~\ref{fig:RC_compare}(a). The rocking curves are measured across the symmetric $30\Bar{3}0$ peak. The full width at half maxima (FWHM) of $\omega$ quantifies the out-of-plane mosaic spread of the \agao\ film. For conventionally grown films the out-of-plane crystal distribution is $\Delta\omega \approx 0.55^{\circ}$ and for MOCATAXY grown films it is $\Delta\omega \approx 0.45^{\circ}$. The film thicknesses $d$ of the conventionally and MOCATAXY grown films are $d = 73\,\text{nm}$ and $d = 127\,\text{nm}$, respectively. Jinno \textit{et al.}, reported that \agao\ films are fully relaxed for $d > 60\,\text{nm}$ \cite{Jinno2021}. Since lattice mismatch and relaxation are not impacted by MOCATAXY, it is noteworthy that despite the MOCATAXY film being thicker, $\Delta\omega$ is substantially smaller compared to what is obtained by conventional growth. The same MOCATAXY grown sample shown here is studied by STEM and shown in Fig.~\ref{fig:4}.

\noindent
Surface morphologies and root mean square roughnesses ($R_{\text{q}}$) are measured by AFM and depicted in Figs.~\ref{fig:RC_compare}(d) and \ref{fig:RC_compare}(e). The best surface roughness for conventionally grown \agao with $d = 66\,\text{nm}$ is $R_\text{q} = 0.64\,\text{nm}$, while the smoothest one for MOCATAXY grown samples with $d \sim 270\,\text{nm}$ has an $R_\text{q} = 0.94\,\text{nm}$. The larger surface roughness for the MOCATAXY grown sample is likely due to facetting on the top surface of the \agao\ thin film [see Fig.~\ref{fig:4}(a)]. We speculate that In does not only act as a catalyst but also acts as a surface active agent (surfactant) for the growth \agao\ thin films. It is widely understood that In can act as a surfactant for the epitaxial growth of GaN-based films \cite{Neugebauer2003a},  and has also been observed during the growth of \ce{\beta-Ga2O3} \cite{Mauze2020} and \ce{\beta-(Al,Ga)2O3} \cite{Vogt2018}. Depending on the growth conditions and growth surface, which can affect the surface diffusion kinetics, surface chemical potentials, and the assessed growth mode, the suppression of facetting may be accomplished through the use of optimized conditions using In as a surfactant, enabling a modification in the surface free energies of the growing \agao\ thin film and a change in its growth mode \cite{Copel1989,Neugebauer2003b,Vogt2018_MOCATAXY,Mauze2020}. However, surfactant-induced morphological phase-transitions from 2-dimensional (2D) layer growth to 3-dimensional (3D) island growth have also been observed during MBE growth \cite{Lewis2017}. We believe that a similar effect occurs for the \ce{\alpha-Ga2O3} surfaces studied here when In may act as an (anti)surfactant during the growth of these films. Note, we have \textit{not} fully explored all  growth regimes made accessible through MOCATAXY in this study. Further studies may lead to additional improvements in the crystalline quality and surface morphologies of the \agao\ thin films. 

\noindent
In Figs.~\ref{fig:allCompare}(a) and \ref{fig:allCompare}(b), the impact of \fga\ and \tg, respectively, on $\Delta\omega$ for samples grown by conventional MBE in the O-rich regime (blue squares) and in the \gr-plateau regime (green circles), as well as for samples grown by MOCATAXY (red stars), are shown. XRD data and $\Delta\omega$ are obtained by the same methods as described above for Fig.~\ref{fig:RC_compare}. Within the O-rich regime at $T_{\text{G}} = 640\,^{\circ}\text{C}$, a large $\Delta\omega$ is observed, Fig.~\ref{fig:allCompare}(a). At higher growth temperatures (i.e. $T_{\text{G}} \geq 660\,^{\circ}\text{C}$), $\Delta\omega$ are similar (or slightly improving) with increasing temperature, regardless of growth regime. We speculate that the reason the crystal quality improves with \tg, is that there is an increase in the kinetic energy and a subsequent increase in the diffusion length of the adsorbates, allowing the Ga and O to reach the proper lattice site. However, if \tg\ is increased too much, a decrease in the surface lifetime of Ga adsorbates may occur, resulting in a reduction in the crystalline quality of the growing thin films. Using MOCATAXY in the Ga-rich regime and fixed \tg, excess Ga may now reduce the needed surface diffusion length, improving the crystalline quality of the obtained \agao\ layers. More studies to separate the effects of \fga\ and \tg\ on $\Delta \omega$ need to be performed, but to the best of our knowledge, the obtained $\Delta \omega$ values are the lowest reported in the literature for \agao\ grown on \aalo.

\noindent
Finally, to directly quantify and identify how MOCATAXY affects the crystal structure of \agao\ thin films, high-angle annular dark-field STEM (HAADF-STEM) was performed along the $<0\bar{1}10>$ zone axis, and is plotted in Fig.~\ref{fig:4}. The sample shown here is the same as the one shown in  Fig.~\ref{fig:RC_compare}(c). In Fig.~\ref{fig:4}(a), a clear contrast differentiates the sapphire substrate, the epitaxial film (\agao), and the protective Au-Pd sputtered coating. The bright contrast observed at the substrate/film interface (see Fig.~\ref{fig:4}(b) and Ref.~\cite{Note1}) is due to additional scattering of the electron beam and indicates the presence of misfit dislocations. These dislocations arise due to the film relaxation caused by strain. A subset of the observed misfit dislocations propagate and lead to threading dislocations. From the contrast variation observed within the film [see Fig.~\ref{fig:4}(a)], an average frequency of one threading dislocation every $30\,\text{nm}$ laterally along the film/substrate interface is observed. While more investigation is needed to determine the cause of the faceting and verify the above hypothesis (e.g., due to the changed growth mode when using In-mediated catalysis), it is observed that the threading dislocations can merge and then continue to propagate toward the film surface. These dislocations terminate at the bottom of intersecting surface planes, where faceting along the ($10\Bar{1}1$) plane is observed. The complimentary facet is unidentified since the facet is not perpendicular to the beam and tilts out of plane. This tilting is detected in Fig.~\ref{fig:4}(a) by the fading of contrast along the surface, in contrast to the sharp change in contrast on the ($10\Bar{1}1$) plane.

\noindent
Figure~\ref{fig:4}(b) shows an enlarged image of the film/substrate interface. A pair of edge dislocations is observed and is highlighted with their Burgers circuits. This edge dislocation pair is observed along the film/substrate interface, and its dislocation density is estimated to be $5 \times 10^{5}\,\text{cm}^{-1}$ (or $ \sim 10^{11}\,\text{cm}^{-2}$), i.e., occurring every 20 nm. This is similar to what is reported by conventional MBE \cite{Jinno2021}. To quantify Al/Ga inter-diffusion at the interface, a line scan (see S-Fig.~2 \cite{Note1}) was performed to quantify the contrast change. An interface width of $\sigma \approx 0.9\,\text{nm}$ was measured from an error function fitted to the Al intensity line scan profile (see S-Fig.~2 \cite{Note1}).

\noindent
A fast Fourier transform (FFT), of the interface region shown in Fig.~\ref{fig:4}(b), is displayed in Fig.~\ref{fig:4}(c). A thin film completely strained to the substrate will show a singular diffraction peak. However, when the film relaxes its interplanar spacing $d_{hkl}$ changes, resulting in an additional peak, shifted from the substrate peak.  However, shifted peaks in the in-plane direction are not visible because the \agao\ ($000\Bar{6}$) reflection peak is approximately 10x weaker than in \aalo. The strain relaxation is observed in the $20\Bar{2}\Bar{4}$ and $10\Bar{1}10$ diffraction peaks of \agao. The strain relaxation is accomplished by the formation of edge dislocations at the interface, where the $20\Bar{2}\Bar{4}$ peak is correlated to the yellow Burgers circuit and the $10\Bar{1}10$ peak to the cyan Burgers circuit. In addition, no phase separation or secondary phases were observed by STEM within the \agao\ film grown by MOCATAXY. However, a bi-layer structure from overlapping \agao\ grains when viewed in projection is observed with a slip along the [$10\Bar{2} \Bar{2}$] direction (see S-Fig.~3 \cite{Note1}). The presence of this bi-layer structure indicates that the film is not single-crystalline. The bi-layer structure was confirmed using an \textit{ab initio} TEM (\textit{ab}TEM) simulation \cite{abtem} which produced a matching HAADF image from the crystallographic information framework. 

\noindent
This TEM investigation of MOCATAXY grown \agao\ shows comparable crystal quality to what is measured for conventional MBE \cite{Jinno2021} with regards to edge dislocation density and phase purity.  We note that the difference in projection direction may have prevented imaging of the bi-layer structure in this previous report. No faceting of \agao\ was observed by conventional MBE when grown on $m$-plane \aalo\ \cite{Jinno2021,mccandless2021}.

\section{Conclusion}
\noindent
Phase-pure \agao($10\Bar{1}0$) on \aalo($10\Bar{1}0$) was grown using conventional MBE and MOCATAXY with thickness up to $d = 262\,\text{nm}$. We mapped out the \gr-dependence on \fga\ and \tg\ and its impact on the crystalline quality and surface morphologies. We identified and explored previously inaccessible growth regimes by MOCATAXY, and showed that it vastly extends the growth regime and improves the out-of-plane mosaic spread of the grown \agao\ films. Using In-mediated catalysis, we observe facetting on top of the \agao($10\Bar{1}0$) layers. This study confirms that this new MBE growth mode can be applied to the growth of \agao -- and is not limited to the growth of the \bgao\ and \balgao\ polymorphs. We emphasize more studies are needed to determine the kinetic parameters that form \agao\ during conventional MBE and MOCATAXY growth, as well as to further improve the quality of the grown \agao/\aalo\ heterostructures, and to understand the mechanisms leading to the surface faceting of \agao.

\section{Acknowledgements}
\noindent
This research is supported by the Air Force Research Laboratory-Cornell Center for Epitaxial Solutions (ACCESS), monitored by Dr. Ali Sayir (FA9550-18-1-0529). JPM acknowledges the support of a National Science Foundation Graduate Research Fellowship under Grant No. DGE–2139899. M. A-O acknowledges financial support from the Central Research Development Fund (CRDF) of the University of Bremen. This work makes use of PARADIM under Cooperative Agreement No. DMR-2039380. This work uses the CCMR and CESI Shared Facilities partly sponsored by the NSF MRSEC program (DMR-1719875) and MRI DMR-1338010, and the Kavli Institute at Cornell (KIC). 

\bibliography{MOCATAXY.bib}

\begin{thebibliography}{33}%
\makeatletter
\providecommand \@ifxundefined [1]{%
 \@ifx{#1\undefined}
}%
\providecommand \@ifnum [1]{%
 \ifnum #1\expandafter \@firstoftwo
 \else \expandafter \@secondoftwo
 \fi
}%
\providecommand \@ifx [1]{%
 \ifx #1\expandafter \@firstoftwo
 \else \expandafter \@secondoftwo
 \fi
}%
\providecommand \natexlab [1]{#1}%
\providecommand \enquote  [1]{``#1''}%
\providecommand \bibnamefont  [1]{#1}%
\providecommand \bibfnamefont [1]{#1}%
\providecommand \citenamefont [1]{#1}%
\providecommand \href@noop [0]{\@secondoftwo}%
\providecommand \href [0]{\begingroup \@sanitize@url \@href}%
\providecommand \@href[1]{\@@startlink{#1}\@@href}%
\providecommand \@@href[1]{\endgroup#1\@@endlink}%
\providecommand \@sanitize@url [0]{\catcode `\\12\catcode `\$12\catcode
  `\&12\catcode `\#12\catcode `\^12\catcode `\_12\catcode `\%12\relax}%
\providecommand \@@startlink[1]{}%
\providecommand \@@endlink[0]{}%
\providecommand \url  [0]{\begingroup\@sanitize@url \@url }%
\providecommand \@url [1]{\endgroup\@href {#1}{\urlprefix }}%
\providecommand \urlprefix  [0]{URL }%
\providecommand \Eprint [0]{\href }%
\providecommand \doibase [0]{https://doi.org/}%
\providecommand \selectlanguage [0]{\@gobble}%
\providecommand \bibinfo  [0]{\@secondoftwo}%
\providecommand \bibfield  [0]{\@secondoftwo}%
\providecommand \translation [1]{[#1]}%
\providecommand \BibitemOpen [0]{}%
\providecommand \bibitemStop [0]{}%
\providecommand \bibitemNoStop [0]{.\EOS\space}%
\providecommand \EOS [0]{\spacefactor3000\relax}%
\providecommand \BibitemShut  [1]{\csname bibitem#1\endcsname}%
\let\auto@bib@innerbib\@empty
\bibitem [{\citenamefont {Tippins}(1965)}]{Tippins1965}%
  \BibitemOpen
  \bibfield  {author} {\bibinfo {author} {\bibfnamefont {H.~H.}\ \bibnamefont
  {Tippins}},\ }\bibfield  {title} {\bibinfo {title} {{Optical absorption and
  photoconductivity in the band edge of \ce{\beta-Ga2O3}}},\ }\bibfield
  {journal} {\bibinfo  {journal} {Physical Review}\ }\textbf {\bibinfo {volume}
  {140}},\ \href {https://doi.org/10.1103/PhysRev.140.A316}
  {10.1103/PhysRev.140.A316} (\bibinfo {year} {1965})\BibitemShut {NoStop}%
\bibitem [{\citenamefont {Galazka}\ \emph {et~al.}(2017)\citenamefont
  {Galazka}, \citenamefont {Uecker}, \citenamefont {Klimm}, \citenamefont
  {Irmscher}, \citenamefont {Naumann}, \citenamefont {Pietsch}, \citenamefont
  {Kwasniewski}, \citenamefont {Bertram}, \citenamefont {Ganschow},\ and\
  \citenamefont {Bickermann}}]{Galazka2017}%
  \BibitemOpen
  \bibfield  {author} {\bibinfo {author} {\bibfnamefont {Z.}~\bibnamefont
  {Galazka}}, \bibinfo {author} {\bibfnamefont {R.}~\bibnamefont {Uecker}},
  \bibinfo {author} {\bibfnamefont {D.}~\bibnamefont {Klimm}}, \bibinfo
  {author} {\bibfnamefont {K.}~\bibnamefont {Irmscher}}, \bibinfo {author}
  {\bibfnamefont {M.}~\bibnamefont {Naumann}}, \bibinfo {author} {\bibfnamefont
  {M.}~\bibnamefont {Pietsch}}, \bibinfo {author} {\bibfnamefont
  {A.}~\bibnamefont {Kwasniewski}}, \bibinfo {author} {\bibfnamefont
  {R.}~\bibnamefont {Bertram}}, \bibinfo {author} {\bibfnamefont
  {S.}~\bibnamefont {Ganschow}},\ and\ \bibinfo {author} {\bibfnamefont
  {M.}~\bibnamefont {Bickermann}},\ }\bibfield  {title} {\bibinfo {title}
  {{Scaling-Up of Bulk \ce{\beta-Ga2O3} Single Crystals by the Czochralski
  Method}},\ }\href {https://doi.org/10.1149/2.0021702jss} {\bibfield
  {journal} {\bibinfo  {journal} {ECS Journal of Solid State Science and
  Technology}\ }\textbf {\bibinfo {volume} {6}},\ \bibinfo {pages} {Q3007}
  (\bibinfo {year} {2017})}\BibitemShut {NoStop}%
\bibitem [{\citenamefont {Kuramata}\ \emph {et~al.}(2016)\citenamefont
  {Kuramata}, \citenamefont {Koshi}, \citenamefont {Watanabe}, \citenamefont
  {Yamaoka}, \citenamefont {Masui},\ and\ \citenamefont
  {Yamakoshi}}]{Kuramata2016_substrates}%
  \BibitemOpen
  \bibfield  {author} {\bibinfo {author} {\bibfnamefont {A.}~\bibnamefont
  {Kuramata}}, \bibinfo {author} {\bibfnamefont {K.}~\bibnamefont {Koshi}},
  \bibinfo {author} {\bibfnamefont {S.}~\bibnamefont {Watanabe}}, \bibinfo
  {author} {\bibfnamefont {Y.}~\bibnamefont {Yamaoka}}, \bibinfo {author}
  {\bibfnamefont {T.}~\bibnamefont {Masui}},\ and\ \bibinfo {author}
  {\bibfnamefont {S.}~\bibnamefont {Yamakoshi}},\ }\bibfield  {title} {\bibinfo
  {title} {{High-quality \ce{\beta-Ga2O3} single crystals grown by edge-defined
  film-fed growth}},\ }\href@noop {} {\bibfield  {journal} {\bibinfo  {journal}
  {Japanese Journal of Applied Physics}\ }\textbf {\bibinfo {volume} {55}}
  (\bibinfo {year} {2016})}\BibitemShut {NoStop}%
\bibitem [{\citenamefont {Bhuiyan}\ \emph {et~al.}(2020)\citenamefont
  {Bhuiyan}, \citenamefont {Feng}, \citenamefont {Johnson}, \citenamefont
  {Huang}, \citenamefont {Sarker}, \citenamefont {Zhu}, \citenamefont {Karim},
  \citenamefont {Mazumder}, \citenamefont {Hwang},\ and\ \citenamefont
  {Zhao}}]{Bhuiyan2020}%
  \BibitemOpen
  \bibfield  {author} {\bibinfo {author} {\bibfnamefont {A.~F.~U.}\
  \bibnamefont {Bhuiyan}}, \bibinfo {author} {\bibfnamefont {Z.}~\bibnamefont
  {Feng}}, \bibinfo {author} {\bibfnamefont {J.~M.}\ \bibnamefont {Johnson}},
  \bibinfo {author} {\bibfnamefont {H.~L.}\ \bibnamefont {Huang}}, \bibinfo
  {author} {\bibfnamefont {J.}~\bibnamefont {Sarker}}, \bibinfo {author}
  {\bibfnamefont {M.}~\bibnamefont {Zhu}}, \bibinfo {author} {\bibfnamefont
  {M.~R.}\ \bibnamefont {Karim}}, \bibinfo {author} {\bibfnamefont
  {B.}~\bibnamefont {Mazumder}}, \bibinfo {author} {\bibfnamefont
  {J.}~\bibnamefont {Hwang}},\ and\ \bibinfo {author} {\bibfnamefont
  {H.}~\bibnamefont {Zhao}},\ }\bibfield  {title} {\bibinfo {title} {{Phase
  transformation in MOCVD growth of $(Al_xGa_{1-x})_2O_3$ thin films}},\ }\href
  {https://doi.org/10.1063/1.5140345} {\bibfield  {journal} {\bibinfo
  {journal} {APL Materials}\ }\textbf {\bibinfo {volume} {8}},\ \bibinfo
  {pages} {031104} (\bibinfo {year} {2020})}\BibitemShut {NoStop}%
\bibitem [{\citenamefont {Jinno}\ \emph {et~al.}(2021)\citenamefont {Jinno},
  \citenamefont {Chang}, \citenamefont {Onuma}, \citenamefont {Cho},
  \citenamefont {Ho}, \citenamefont {Rowe}, \citenamefont {Cao}, \citenamefont
  {Lee}, \citenamefont {Protasenko}, \citenamefont {Schlom}, \citenamefont
  {Muller}, \citenamefont {Xing},\ and\ \citenamefont {Jena}}]{Jinno2021}%
  \BibitemOpen
  \bibfield  {author} {\bibinfo {author} {\bibfnamefont {R.}~\bibnamefont
  {Jinno}}, \bibinfo {author} {\bibfnamefont {C.~S.}\ \bibnamefont {Chang}},
  \bibinfo {author} {\bibfnamefont {T.}~\bibnamefont {Onuma}}, \bibinfo
  {author} {\bibfnamefont {Y.}~\bibnamefont {Cho}}, \bibinfo {author}
  {\bibfnamefont {S.~T.}\ \bibnamefont {Ho}}, \bibinfo {author} {\bibfnamefont
  {D.}~\bibnamefont {Rowe}}, \bibinfo {author} {\bibfnamefont {M.~C.}\
  \bibnamefont {Cao}}, \bibinfo {author} {\bibfnamefont {K.}~\bibnamefont
  {Lee}}, \bibinfo {author} {\bibfnamefont {V.}~\bibnamefont {Protasenko}},
  \bibinfo {author} {\bibfnamefont {D.~G.}\ \bibnamefont {Schlom}}, \bibinfo
  {author} {\bibfnamefont {D.~A.}\ \bibnamefont {Muller}}, \bibinfo {author}
  {\bibfnamefont {H.~G.}\ \bibnamefont {Xing}},\ and\ \bibinfo {author}
  {\bibfnamefont {D.}~\bibnamefont {Jena}},\ }\bibfield  {title} {\bibinfo
  {title} {{Crystal orientation dictated epitaxy of ultrawide-bandgap 5.4-to
  8.6-eV $\alpha -(AlGa)_2O_3$ on m-plane sapphire}},\ }\href
  {https://doi.org/10.1126/sciadv.abd5891} {\bibfield  {journal} {\bibinfo
  {journal} {Science Advances}\ }\textbf {\bibinfo {volume} {7}},\ \bibinfo
  {pages} {1} (\bibinfo {year} {2021})},\ \Eprint
  {https://arxiv.org/abs/2007.03415} {arXiv:2007.03415} \BibitemShut {NoStop}%
\bibitem [{\citenamefont {Bhuiyan}\ \emph {et~al.}(2021)\citenamefont
  {Bhuiyan}, \citenamefont {Feng}, \citenamefont {Huang}, \citenamefont {Meng},
  \citenamefont {Hwang},\ and\ \citenamefont {Zhao}}]{Bhuiyan2021_alphGO}%
  \BibitemOpen
  \bibfield  {author} {\bibinfo {author} {\bibfnamefont {A.~F.~U.}\
  \bibnamefont {Bhuiyan}}, \bibinfo {author} {\bibfnamefont {Z.}~\bibnamefont
  {Feng}}, \bibinfo {author} {\bibfnamefont {H.~L.}\ \bibnamefont {Huang}},
  \bibinfo {author} {\bibfnamefont {L.}~\bibnamefont {Meng}}, \bibinfo {author}
  {\bibfnamefont {J.}~\bibnamefont {Hwang}},\ and\ \bibinfo {author}
  {\bibfnamefont {H.}~\bibnamefont {Zhao}},\ }\bibfield  {title} {\bibinfo
  {title} {{Metalorganic chemical vapor deposition of \ce{\alpha-Ga2O3} and
  \ce{\alpha-Ga2O3} thin films on m-plane sapphire substrates}},\ }\bibfield
  {journal} {\bibinfo  {journal} {APL Materials}\ }\textbf {\bibinfo {volume}
  {9}},\ \href {https://doi.org/10.1063/5.0065087} {10.1063/5.0065087}
  (\bibinfo {year} {2021})\BibitemShut {NoStop}%
\bibitem [{\citenamefont {Tsai}\ \emph {et~al.}(2010)\citenamefont {Tsai},
  \citenamefont {Bierwagen}, \citenamefont {White},\ and\ \citenamefont
  {Speck}}]{Tsai2009}%
  \BibitemOpen
  \bibfield  {author} {\bibinfo {author} {\bibfnamefont {M.-Y.}\ \bibnamefont
  {Tsai}}, \bibinfo {author} {\bibfnamefont {O.}~\bibnamefont {Bierwagen}},
  \bibinfo {author} {\bibfnamefont {M.~E.}\ \bibnamefont {White}},\ and\
  \bibinfo {author} {\bibfnamefont {J.~S.}\ \bibnamefont {Speck}},\ }\bibfield
  {title} {\bibinfo {title} {{\ce{\beta-Ga2O3} growth by plasma-assisted
  molecular beam epitaxy}},\ }\href {https://doi.org/10.1116/1.3294715}
  {\bibfield  {journal} {\bibinfo  {journal} {Journal of Vacuum Science and
  Technology A: Vacuum, Surfaces, and Films}\ }\textbf {\bibinfo {volume}
  {28}},\ \bibinfo {pages} {354} (\bibinfo {year} {2010})}\BibitemShut
  {NoStop}%
\bibitem [{\citenamefont {Cassabois}\ \emph {et~al.}(2016)\citenamefont
  {Cassabois}, \citenamefont {Valvin},\ and\ \citenamefont
  {Gil}}]{Cassabois2016}%
  \BibitemOpen
  \bibfield  {author} {\bibinfo {author} {\bibfnamefont {G.}~\bibnamefont
  {Cassabois}}, \bibinfo {author} {\bibfnamefont {P.}~\bibnamefont {Valvin}},\
  and\ \bibinfo {author} {\bibfnamefont {B.}~\bibnamefont {Gil}},\ }\bibfield
  {title} {\bibinfo {title} {{Hexagonal boron nitride is an indirect bandgap
  semiconductor}},\ }\href {https://doi.org/10.1038/nphoton.2015.277}
  {\bibfield  {journal} {\bibinfo  {journal} {Nature Photonics}\ }\textbf
  {\bibinfo {volume} {10}},\ \bibinfo {pages} {262} (\bibinfo {year} {2016})},\
  \Eprint {https://arxiv.org/abs/1512.02962} {arXiv:1512.02962} \BibitemShut
  {NoStop}%
\bibitem [{\citenamefont {McCandless}\ \emph {et~al.}(2021)\citenamefont
  {McCandless}, \citenamefont {Chang}, \citenamefont {Nomoto}, \citenamefont
  {Casamento}, \citenamefont {Protasenko}, \citenamefont {Vogt}, \citenamefont
  {Rowe}, \citenamefont {Gann}, \citenamefont {Ho}, \citenamefont {Li},
  \citenamefont {Jinno}, \citenamefont {Cho}, \citenamefont {Green},
  \citenamefont {Chabak}, \citenamefont {Schlom}, \citenamefont {Thompson},
  \citenamefont {Muller}, \citenamefont {Xing},\ and\ \citenamefont
  {Jena}}]{mccandless2021}%
  \BibitemOpen
  \bibfield  {author} {\bibinfo {author} {\bibfnamefont {J.~P.}\ \bibnamefont
  {McCandless}}, \bibinfo {author} {\bibfnamefont {C.~S.}\ \bibnamefont
  {Chang}}, \bibinfo {author} {\bibfnamefont {K.}~\bibnamefont {Nomoto}},
  \bibinfo {author} {\bibfnamefont {J.}~\bibnamefont {Casamento}}, \bibinfo
  {author} {\bibfnamefont {V.}~\bibnamefont {Protasenko}}, \bibinfo {author}
  {\bibfnamefont {P.}~\bibnamefont {Vogt}}, \bibinfo {author} {\bibfnamefont
  {D.}~\bibnamefont {Rowe}}, \bibinfo {author} {\bibfnamefont {K.}~\bibnamefont
  {Gann}}, \bibinfo {author} {\bibfnamefont {S.~T.}\ \bibnamefont {Ho}},
  \bibinfo {author} {\bibfnamefont {W.}~\bibnamefont {Li}}, \bibinfo {author}
  {\bibfnamefont {R.}~\bibnamefont {Jinno}}, \bibinfo {author} {\bibfnamefont
  {Y.}~\bibnamefont {Cho}}, \bibinfo {author} {\bibfnamefont {A.~J.}\
  \bibnamefont {Green}}, \bibinfo {author} {\bibfnamefont {K.~D.}\ \bibnamefont
  {Chabak}}, \bibinfo {author} {\bibfnamefont {D.~G.}\ \bibnamefont {Schlom}},
  \bibinfo {author} {\bibfnamefont {M.~O.}\ \bibnamefont {Thompson}}, \bibinfo
  {author} {\bibfnamefont {D.~A.}\ \bibnamefont {Muller}}, \bibinfo {author}
  {\bibfnamefont {H.~G.}\ \bibnamefont {Xing}},\ and\ \bibinfo {author}
  {\bibfnamefont {D.}~\bibnamefont {Jena}},\ }\bibfield  {title} {\bibinfo
  {title} {{Thermal stability of epitaxial \ce{\alpha-Ga2O3} and
  \ce{(Al,Ga)2O3} layers on m-plane sapphire}},\ }\href
  {https://doi.org/10.1063/5.0064278} {\bibfield  {journal} {\bibinfo
  {journal} {Applied Physics Letters}\ }\textbf {\bibinfo {volume} {119}},\
  \bibinfo {pages} {062101} (\bibinfo {year} {2021})}\BibitemShut {NoStop}%
\bibitem [{\citenamefont {Akaiwa}\ and\ \citenamefont
  {Fujita}(2012)}]{Akaiwa2012}%
  \BibitemOpen
  \bibfield  {author} {\bibinfo {author} {\bibfnamefont {K.}~\bibnamefont
  {Akaiwa}}\ and\ \bibinfo {author} {\bibfnamefont {S.}~\bibnamefont
  {Fujita}},\ }\bibfield  {title} {\bibinfo {title} {{Electrical conductive
  corundum-structured \ce{\alpha-Ga2O3} Thin films on sapphire with tin-doping
  grown by spray-assisted mist chemical vapor deposition}},\ }\href
  {https://doi.org/10.1143/JJAP.51.070203} {\bibfield  {journal} {\bibinfo
  {journal} {Japanese Journal of Applied Physics}\ }\textbf {\bibinfo {volume}
  {51}},\ \bibinfo {pages} {070203} (\bibinfo {year} {2012})}\BibitemShut
  {NoStop}%
\bibitem [{\citenamefont {Uchida}\ \emph {et~al.}(2018)\citenamefont {Uchida},
  \citenamefont {Kaneko},\ and\ \citenamefont {Fujita}}]{uchida2018}%
  \BibitemOpen
  \bibfield  {author} {\bibinfo {author} {\bibfnamefont {T.}~\bibnamefont
  {Uchida}}, \bibinfo {author} {\bibfnamefont {K.}~\bibnamefont {Kaneko}},\
  and\ \bibinfo {author} {\bibfnamefont {S.}~\bibnamefont {Fujita}},\
  }\bibfield  {title} {\bibinfo {title} {{Electrical characterization of
  Si-doped n-type \ce{\alpha-Ga2O3} on sapphire substrates}},\ }\href
  {https://doi.org/10.1557/adv.2018.45} {\bibfield  {journal} {\bibinfo
  {journal} {MRS Advances}\ }\textbf {\bibinfo {volume} {3}},\ \bibinfo {pages}
  {171} (\bibinfo {year} {2018})}\BibitemShut {NoStop}%
\bibitem [{\citenamefont {Varley}\ \emph {et~al.}(2010)\citenamefont {Varley},
  \citenamefont {Weber}, \citenamefont {Janotti},\ and\ \citenamefont {{Van De
  Walle}}}]{Varley2010}%
  \BibitemOpen
  \bibfield  {author} {\bibinfo {author} {\bibfnamefont {J.~B.}\ \bibnamefont
  {Varley}}, \bibinfo {author} {\bibfnamefont {J.~R.}\ \bibnamefont {Weber}},
  \bibinfo {author} {\bibfnamefont {A.}~\bibnamefont {Janotti}},\ and\ \bibinfo
  {author} {\bibfnamefont {C.~G.}\ \bibnamefont {{Van De Walle}}},\ }\bibfield
  {title} {\bibinfo {title} {{Oxygen vacancies and donor impurities in
  \ce{\beta-Ga2O3}}},\ }\href {https://doi.org/10.1063/1.3499306} {\bibfield
  {journal} {\bibinfo  {journal} {Applied Physics Letters}\ }\textbf {\bibinfo
  {volume} {97}},\ \bibinfo {pages} {97} (\bibinfo {year} {2010})}\BibitemShut
  {NoStop}%
\bibitem [{\citenamefont {Varley}\ \emph {et~al.}(2011)\citenamefont {Varley},
  \citenamefont {Peelaers}, \citenamefont {Janotti},\ and\ \citenamefont {{Van
  De Walle}}}]{Varley2011}%
  \BibitemOpen
  \bibfield  {author} {\bibinfo {author} {\bibfnamefont {J.~B.}\ \bibnamefont
  {Varley}}, \bibinfo {author} {\bibfnamefont {H.}~\bibnamefont {Peelaers}},
  \bibinfo {author} {\bibfnamefont {A.}~\bibnamefont {Janotti}},\ and\ \bibinfo
  {author} {\bibfnamefont {C.~G.}\ \bibnamefont {{Van De Walle}}},\ }\bibfield
  {title} {\bibinfo {title} {{Hydrogenated cation vacancies in semiconducting
  oxides}},\ }\bibfield  {journal} {\bibinfo  {journal} {Journal of Physics:
  Condensed MatterCondensed Matter}\ }\textbf {\bibinfo {volume} {23}},\ \href
  {https://doi.org/10.1088/0953-8984/23/33/334212}
  {10.1088/0953-8984/23/33/334212} (\bibinfo {year} {2011})\BibitemShut
  {NoStop}%
\bibitem [{\citenamefont {Rafique}\ \emph {et~al.}(2018)\citenamefont
  {Rafique}, \citenamefont {Han}, \citenamefont {Neal}, \citenamefont {Mou},
  \citenamefont {Boeckl},\ and\ \citenamefont {Zhao}}]{Rafique2018}%
  \BibitemOpen
  \bibfield  {author} {\bibinfo {author} {\bibfnamefont {S.}~\bibnamefont
  {Rafique}}, \bibinfo {author} {\bibfnamefont {L.}~\bibnamefont {Han}},
  \bibinfo {author} {\bibfnamefont {A.~T.}\ \bibnamefont {Neal}}, \bibinfo
  {author} {\bibfnamefont {S.}~\bibnamefont {Mou}}, \bibinfo {author}
  {\bibfnamefont {J.}~\bibnamefont {Boeckl}},\ and\ \bibinfo {author}
  {\bibfnamefont {H.}~\bibnamefont {Zhao}},\ }\bibfield  {title} {\bibinfo
  {title} {{Towards High-Mobility Heteroepitaxial \ce{\beta-Ga2O3} on Sapphire
  Dependence on The Substrate Off Axis Angle}},\ }\href
  {https://doi.org/10.1002/pssa.201700467} {\bibfield  {journal} {\bibinfo
  {journal} {Physica Status Solidi (A) Applications and Materials Science}\
  }\textbf {\bibinfo {volume} {215}},\ \bibinfo {pages} {1700467} (\bibinfo
  {year} {2018})}\BibitemShut {NoStop}%
\bibitem [{\citenamefont {Korhonen}\ \emph {et~al.}(2015)\citenamefont
  {Korhonen}, \citenamefont {Tuomisto}, \citenamefont {Gogova}, \citenamefont
  {Wagner}, \citenamefont {Baldini}, \citenamefont {Galazka}, \citenamefont
  {Schewski},\ and\ \citenamefont {Albrecht}}]{Korhonen2015}%
  \BibitemOpen
  \bibfield  {author} {\bibinfo {author} {\bibfnamefont {E.}~\bibnamefont
  {Korhonen}}, \bibinfo {author} {\bibfnamefont {F.}~\bibnamefont {Tuomisto}},
  \bibinfo {author} {\bibfnamefont {D.}~\bibnamefont {Gogova}}, \bibinfo
  {author} {\bibfnamefont {G.}~\bibnamefont {Wagner}}, \bibinfo {author}
  {\bibfnamefont {M.}~\bibnamefont {Baldini}}, \bibinfo {author} {\bibfnamefont
  {Z.}~\bibnamefont {Galazka}}, \bibinfo {author} {\bibfnamefont
  {R.}~\bibnamefont {Schewski}},\ and\ \bibinfo {author} {\bibfnamefont
  {M.}~\bibnamefont {Albrecht}},\ }\bibfield  {title} {\bibinfo {title}
  {{Electrical compensation by Ga vacancies in \ce{Ga2O3} thin films}},\ }\href
  {https://doi.org/10.1063/1.4922814} {\bibfield  {journal} {\bibinfo
  {journal} {Applied Physics Letters}\ }\textbf {\bibinfo {volume} {106}},\
  \bibinfo {pages} {1} (\bibinfo {year} {2015})}\BibitemShut {NoStop}%
\bibitem [{\citenamefont {Vogt}\ \emph {et~al.}(2022)\citenamefont {Vogt},
  \citenamefont {Hensling}, \citenamefont {Azizie}, \citenamefont {McCandless},
  \citenamefont {Park}, \citenamefont {DeLello}, \citenamefont {Muller},
  \citenamefont {Xing}, \citenamefont {Jena},\ and\ \citenamefont
  {Schlom}}]{Vogt2022_PhysRevApp}%
  \BibitemOpen
  \bibfield  {author} {\bibinfo {author} {\bibfnamefont {P.}~\bibnamefont
  {Vogt}}, \bibinfo {author} {\bibfnamefont {F.~V.}\ \bibnamefont {Hensling}},
  \bibinfo {author} {\bibfnamefont {K.}~\bibnamefont {Azizie}}, \bibinfo
  {author} {\bibfnamefont {J.~P.}\ \bibnamefont {McCandless}}, \bibinfo
  {author} {\bibfnamefont {J.}~\bibnamefont {Park}}, \bibinfo {author}
  {\bibfnamefont {K.}~\bibnamefont {DeLello}}, \bibinfo {author} {\bibfnamefont
  {D.~A.}\ \bibnamefont {Muller}}, \bibinfo {author} {\bibfnamefont {H.~G.}\
  \bibnamefont {Xing}}, \bibinfo {author} {\bibfnamefont {D.}~\bibnamefont
  {Jena}},\ and\ \bibinfo {author} {\bibfnamefont {D.~G.}\ \bibnamefont
  {Schlom}},\ }\bibfield  {title} {\bibinfo {title} {{Extending the Kinetic and
  Thermodynamic Limits of Molecular-Beam Epitaxy Utilizing Suboxide Sources or
  Metal-Oxide-Catalyzed Epitaxy}},\ }\href
  {https://doi.org/10.1103/physrevapplied.17.034021} {\bibfield  {journal}
  {\bibinfo  {journal} {Physical Review Applied}\ }\textbf {\bibinfo {volume}
  {17}},\ \bibinfo {pages} {034021} (\bibinfo {year} {2022})}\BibitemShut
  {NoStop}%
\bibitem [{\citenamefont {Vogt}\ \emph
  {et~al.}(2018{\natexlab{a}})\citenamefont {Vogt}, \citenamefont {Mauze},
  \citenamefont {Wu}, \citenamefont {Bonef},\ and\ \citenamefont
  {Speck}}]{Vogt2018_MOCATAXY}%
  \BibitemOpen
  \bibfield  {author} {\bibinfo {author} {\bibfnamefont {P.}~\bibnamefont
  {Vogt}}, \bibinfo {author} {\bibfnamefont {A.}~\bibnamefont {Mauze}},
  \bibinfo {author} {\bibfnamefont {F.}~\bibnamefont {Wu}}, \bibinfo {author}
  {\bibfnamefont {B.}~\bibnamefont {Bonef}},\ and\ \bibinfo {author}
  {\bibfnamefont {J.~S.}\ \bibnamefont {Speck}},\ }\bibfield  {title} {\bibinfo
  {title} {{Metal-oxide catalyzed epitaxy (MOCATAXY): The example of the O
  plasma-assisted molecular beam epitaxy of $\beta-(Al_xGa_{1-x}
  )_2O_3$/$\beta-Ga_2O_3$ heterostructures}},\ }\href
  {https://doi.org/10.7567/APEX.11.115503} {\bibfield  {journal} {\bibinfo
  {journal} {Applied Physics Express}\ }\textbf {\bibinfo {volume} {11}},\
  \bibinfo {pages} {1} (\bibinfo {year} {2018}{\natexlab{a}})}\BibitemShut
  {NoStop}%
\bibitem [{\citenamefont {Vogt}\ \emph {et~al.}(2017)\citenamefont {Vogt},
  \citenamefont {Brandt}, \citenamefont {Riechert}, \citenamefont
  {L{\"{a}}hnemann},\ and\ \citenamefont {Bierwagen}}]{Vogt2017_mocataxy}%
  \BibitemOpen
  \bibfield  {author} {\bibinfo {author} {\bibfnamefont {P.}~\bibnamefont
  {Vogt}}, \bibinfo {author} {\bibfnamefont {O.}~\bibnamefont {Brandt}},
  \bibinfo {author} {\bibfnamefont {H.}~\bibnamefont {Riechert}}, \bibinfo
  {author} {\bibfnamefont {J.}~\bibnamefont {L{\"{a}}hnemann}},\ and\ \bibinfo
  {author} {\bibfnamefont {O.}~\bibnamefont {Bierwagen}},\ }\bibfield  {title}
  {\bibinfo {title} {{Metal-Exchange Catalysis in the Growth of Sesquioxides:
  Towards Heterostructures of Transparent Oxide Semiconductors}},\ }\href
  {https://doi.org/10.1103/PhysRevLett.119.196001} {\bibfield  {journal}
  {\bibinfo  {journal} {Physical Review Letters}\ }\textbf {\bibinfo {volume}
  {119}},\ \bibinfo {pages} {6} (\bibinfo {year} {2017})}\BibitemShut {NoStop}%
\bibitem [{\citenamefont {Mazzolini}\ \emph {et~al.}(2019)\citenamefont
  {Mazzolini}, \citenamefont {Vogt}, \citenamefont {Schewski}, \citenamefont
  {Wouters}, \citenamefont {Albrecht},\ and\ \citenamefont
  {Bierwagen}}]{Mazzolini2019}%
  \BibitemOpen
  \bibfield  {author} {\bibinfo {author} {\bibfnamefont {P.}~\bibnamefont
  {Mazzolini}}, \bibinfo {author} {\bibfnamefont {P.}~\bibnamefont {Vogt}},
  \bibinfo {author} {\bibfnamefont {R.}~\bibnamefont {Schewski}}, \bibinfo
  {author} {\bibfnamefont {C.}~\bibnamefont {Wouters}}, \bibinfo {author}
  {\bibfnamefont {M.}~\bibnamefont {Albrecht}},\ and\ \bibinfo {author}
  {\bibfnamefont {O.}~\bibnamefont {Bierwagen}},\ }\bibfield  {title} {\bibinfo
  {title} {{Faceting and metal-exchange catalysis in (010) \ce{\beta-Ga2O3}
  thin films homoepitaxially grown by plasma-assisted molecular beam
  epitaxy}},\ }\href {https://doi.org/10.1063/1.5054386} {\bibfield  {journal}
  {\bibinfo  {journal} {APL Materials}\ }\textbf {\bibinfo {volume} {7}},\
  \bibinfo {pages} {022511} (\bibinfo {year} {2019})}\BibitemShut {NoStop}%
\bibitem [{\citenamefont {Mauze}\ \emph {et~al.}(2020)\citenamefont {Mauze},
  \citenamefont {Zhang}, \citenamefont {Itoh}, \citenamefont {Wu},\ and\
  \citenamefont {Speck}}]{Mauze2020}%
  \BibitemOpen
  \bibfield  {author} {\bibinfo {author} {\bibfnamefont {A.}~\bibnamefont
  {Mauze}}, \bibinfo {author} {\bibfnamefont {Y.}~\bibnamefont {Zhang}},
  \bibinfo {author} {\bibfnamefont {T.}~\bibnamefont {Itoh}}, \bibinfo {author}
  {\bibfnamefont {F.}~\bibnamefont {Wu}},\ and\ \bibinfo {author}
  {\bibfnamefont {J.~S.}\ \bibnamefont {Speck}},\ }\bibfield  {title} {\bibinfo
  {title} {{Metal oxide catalyzed epitaxy (MOCATAXY) of \ce{\beta-Ga2O3} films
  in various orientations grown by plasma-assisted molecular beam epitaxy}},\
  }\href {https://doi.org/10.1063/1.5135930} {\bibfield  {journal} {\bibinfo
  {journal} {APL Materials}\ }\textbf {\bibinfo {volume} {8}},\ \bibinfo
  {pages} {021104} (\bibinfo {year} {2020})}\BibitemShut {NoStop}%
\bibitem [{\citenamefont {Mazzolini}\ \emph {et~al.}(2020)\citenamefont
  {Mazzolini}, \citenamefont {Falkenstein}, \citenamefont {Wouters},
  \citenamefont {Schewski}, \citenamefont {Markurt}, \citenamefont {Galazka},
  \citenamefont {Martin}, \citenamefont {Albrecht},\ and\ \citenamefont
  {Bierwagen}}]{Mazzolini2020}%
  \BibitemOpen
  \bibfield  {author} {\bibinfo {author} {\bibfnamefont {P.}~\bibnamefont
  {Mazzolini}}, \bibinfo {author} {\bibfnamefont {A.}~\bibnamefont
  {Falkenstein}}, \bibinfo {author} {\bibfnamefont {C.}~\bibnamefont
  {Wouters}}, \bibinfo {author} {\bibfnamefont {R.}~\bibnamefont {Schewski}},
  \bibinfo {author} {\bibfnamefont {T.}~\bibnamefont {Markurt}}, \bibinfo
  {author} {\bibfnamefont {Z.}~\bibnamefont {Galazka}}, \bibinfo {author}
  {\bibfnamefont {M.}~\bibnamefont {Martin}}, \bibinfo {author} {\bibfnamefont
  {M.}~\bibnamefont {Albrecht}},\ and\ \bibinfo {author} {\bibfnamefont
  {O.}~\bibnamefont {Bierwagen}},\ }\bibfield  {title} {\bibinfo {title}
  {{Substrate-orientation dependence of \ce{\beta-Ga2O3} (100), (010), (001),
  and ($\bar{2}$01) homoepitaxy by indium-mediated metal-exchange catalyzed
  molecular beam epitaxy (MEXCAT-MBE)}},\ }\href
  {https://doi.org/10.1063/1.5135772} {\bibfield  {journal} {\bibinfo
  {journal} {APL Materials}\ }\textbf {\bibinfo {volume} {8}},\ \bibinfo
  {pages} {011107} (\bibinfo {year} {2020})}\BibitemShut {NoStop}%
\bibitem [{\citenamefont {Kuang}\ \emph {et~al.}(2021)\citenamefont {Kuang},
  \citenamefont {Chen}, \citenamefont {Ma}, \citenamefont {Du}, \citenamefont
  {Zhang}, \citenamefont {Hao}, \citenamefont {Ren}, \citenamefont {Liu},
  \citenamefont {Zhu}, \citenamefont {Gu}, \citenamefont {Zhang}, \citenamefont
  {Zheng},\ and\ \citenamefont {Ye}}]{Kuang2021}%
  \BibitemOpen
  \bibfield  {author} {\bibinfo {author} {\bibfnamefont {Y.}~\bibnamefont
  {Kuang}}, \bibinfo {author} {\bibfnamefont {X.}~\bibnamefont {Chen}},
  \bibinfo {author} {\bibfnamefont {T.}~\bibnamefont {Ma}}, \bibinfo {author}
  {\bibfnamefont {Q.}~\bibnamefont {Du}}, \bibinfo {author} {\bibfnamefont
  {Y.}~\bibnamefont {Zhang}}, \bibinfo {author} {\bibfnamefont
  {J.}~\bibnamefont {Hao}}, \bibinfo {author} {\bibfnamefont {F.~F.}\
  \bibnamefont {Ren}}, \bibinfo {author} {\bibfnamefont {B.}~\bibnamefont
  {Liu}}, \bibinfo {author} {\bibfnamefont {S.}~\bibnamefont {Zhu}}, \bibinfo
  {author} {\bibfnamefont {S.}~\bibnamefont {Gu}}, \bibinfo {author}
  {\bibfnamefont {R.}~\bibnamefont {Zhang}}, \bibinfo {author} {\bibfnamefont
  {Y.}~\bibnamefont {Zheng}},\ and\ \bibinfo {author} {\bibfnamefont
  {J.}~\bibnamefont {Ye}},\ }\bibfield  {title} {\bibinfo {title} {{Band
  Alignment and Enhanced Interfacial Conductivity Manipulated by Polarization
  in a Surfactant-Mediated Grown \ce{\kappa-Ga2O3}/\ce{In2O3}
  Heterostructure}},\ }\href {https://doi.org/10.1021/acsaelm.0c00947}
  {\bibfield  {journal} {\bibinfo  {journal} {ACS Applied Electronic
  Materials}\ }\textbf {\bibinfo {volume} {3}},\ \bibinfo {pages} {795}
  (\bibinfo {year} {2021})}\BibitemShut {NoStop}%
\bibitem [{\citenamefont {Kracht}\ \emph {et~al.}(2017)\citenamefont {Kracht},
  \citenamefont {Karg}, \citenamefont {Sch{\"{o}}rmann}, \citenamefont
  {Weinhold}, \citenamefont {Zink}, \citenamefont {Michel}, \citenamefont
  {Rohnke}, \citenamefont {Schowalter}, \citenamefont {Gerken}, \citenamefont
  {Rosenauer}, \citenamefont {Klar}, \citenamefont {Janek},\ and\ \citenamefont
  {Eickhoff}}]{Kracht2017}%
  \BibitemOpen
  \bibfield  {author} {\bibinfo {author} {\bibfnamefont {M.}~\bibnamefont
  {Kracht}}, \bibinfo {author} {\bibfnamefont {A.}~\bibnamefont {Karg}},
  \bibinfo {author} {\bibfnamefont {J.}~\bibnamefont {Sch{\"{o}}rmann}},
  \bibinfo {author} {\bibfnamefont {M.}~\bibnamefont {Weinhold}}, \bibinfo
  {author} {\bibfnamefont {D.}~\bibnamefont {Zink}}, \bibinfo {author}
  {\bibfnamefont {F.}~\bibnamefont {Michel}}, \bibinfo {author} {\bibfnamefont
  {M.}~\bibnamefont {Rohnke}}, \bibinfo {author} {\bibfnamefont
  {M.}~\bibnamefont {Schowalter}}, \bibinfo {author} {\bibfnamefont
  {B.}~\bibnamefont {Gerken}}, \bibinfo {author} {\bibfnamefont
  {A.}~\bibnamefont {Rosenauer}}, \bibinfo {author} {\bibfnamefont {P.~J.}\
  \bibnamefont {Klar}}, \bibinfo {author} {\bibfnamefont {J.}~\bibnamefont
  {Janek}},\ and\ \bibinfo {author} {\bibfnamefont {M.}~\bibnamefont
  {Eickhoff}},\ }\bibfield  {title} {\bibinfo {title} {{Tin-Assisted Synthesis
  of \ce{\epsilon-Ga2O3} by Molecular Beam Epitaxy}},\ }\href
  {https://doi.org/10.1103/PhysRevApplied.8.054002} {\bibfield  {journal}
  {\bibinfo  {journal} {Physical Review Applied}\ }\textbf {\bibinfo {volume}
  {8}},\ \bibinfo {pages} {054002} (\bibinfo {year} {2017})}\BibitemShut
  {NoStop}%
\bibitem [{\citenamefont {Vogt}(2017)}]{Vogt2017_dissertation}%
  \BibitemOpen
  \bibfield  {author} {\bibinfo {author} {\bibfnamefont {P.}~\bibnamefont
  {Vogt}},\ }\emph {\bibinfo {title} {{Growth Kinetics , Thermodynamics , and
  Phase Formation of group-III and IV oxides during Molecular Beam Epitaxy}}},\
  \href@noop {} {Ph.D. thesis} (\bibinfo {year} {2017})\BibitemShut {NoStop}%
\bibitem [{\citenamefont {Vogt}\ and\ \citenamefont
  {Bierwagen}(2018)}]{vogt2018_model}%
  \BibitemOpen
  \bibfield  {author} {\bibinfo {author} {\bibfnamefont {P.}~\bibnamefont
  {Vogt}}\ and\ \bibinfo {author} {\bibfnamefont {O.}~\bibnamefont
  {Bierwagen}},\ }\bibfield  {title} {\bibinfo {title} {{Quantitative
  subcompound-mediated reaction model for the molecular beam epitaxy of III-VI
  and IV-VI thin films: Applied to \ce{Ga2O3} ,\ce{In2O3}, and \ce{SnO2}}},\
  }\href {https://doi.org/10.1103/PhysRevMaterials.2.120401} {\bibfield
  {journal} {\bibinfo  {journal} {Physical Review Materials}\ }\textbf
  {\bibinfo {volume} {2}},\ \bibinfo {pages} {1} (\bibinfo {year}
  {2018})}\BibitemShut {NoStop}%
\bibitem [{Note1()}]{Note1}%
  \BibitemOpen
  \bibinfo {note} {See Supplemental Material at [URL will be inserted by
  publisher] for a model of $\varGamma $\ as a function of $\bgroup \oldphi
  \egroup _{\protect \text {O}}$\ at $T_{\protect \text {G}}$\ = 680$\protect
  \,^{\circ }\protect \text {C}$\ (S-Fig.~1) as well as images obtained by STEM
  (S-Fig.~2 and S-Fig.~3) and an included crystallographic model.}\BibitemShut
  {Stop}%
\bibitem [{\citenamefont {Vogt}\ and\ \citenamefont
  {Bierwagen}(2016)}]{Vogt2016}%
  \BibitemOpen
  \bibfield  {author} {\bibinfo {author} {\bibfnamefont {P.}~\bibnamefont
  {Vogt}}\ and\ \bibinfo {author} {\bibfnamefont {O.}~\bibnamefont
  {Bierwagen}},\ }\bibfield  {title} {\bibinfo {title} {{Reaction kinetics and
  growth window for plasma-assisted molecular beam epitaxy of $Ga_2O_3$:
  Incorporation of Ga vs. $Ga_2O$ desorption}},\ }\href
  {https://doi.org/10.1063/1.4942002} {\bibfield  {journal} {\bibinfo
  {journal} {Applied Physics Letters}\ }\textbf {\bibinfo {volume} {108}},\
  \bibinfo {pages} {072101} (\bibinfo {year} {2016})}\BibitemShut {NoStop}%
\bibitem [{\citenamefont {Neugebauer}\ \emph {et~al.}(2003)\citenamefont
  {Neugebauer}, \citenamefont {Zywietz}, \citenamefont {Scheffler},
  \citenamefont {Northrup}, \citenamefont {Chen},\ and\ \citenamefont
  {Feenstra}}]{Neugebauer2003a}%
  \BibitemOpen
  \bibfield  {author} {\bibinfo {author} {\bibfnamefont {J.}~\bibnamefont
  {Neugebauer}}, \bibinfo {author} {\bibfnamefont {T.~K.}\ \bibnamefont
  {Zywietz}}, \bibinfo {author} {\bibfnamefont {M.}~\bibnamefont {Scheffler}},
  \bibinfo {author} {\bibfnamefont {J.~E.}\ \bibnamefont {Northrup}}, \bibinfo
  {author} {\bibfnamefont {H.}~\bibnamefont {Chen}},\ and\ \bibinfo {author}
  {\bibfnamefont {R.~M.}\ \bibnamefont {Feenstra}},\ }\bibfield  {title}
  {\bibinfo {title} {{Adatom Kinetics On and Below the Surface: The Existence
  of a New Diffusion Channel}},\ }\href
  {https://doi.org/10.1103/PhysRevLett.90.056101} {\bibfield  {journal}
  {\bibinfo  {journal} {Physical Review Letters}\ }\textbf {\bibinfo {volume}
  {90}},\ \bibinfo {pages} {1} (\bibinfo {year} {2003})}\BibitemShut {NoStop}%
\bibitem [{\citenamefont {Vogt}\ \emph
  {et~al.}(2018{\natexlab{b}})\citenamefont {Vogt}, \citenamefont {Mauze},
  \citenamefont {Wu}, \citenamefont {Bonef},\ and\ \citenamefont
  {Speck}}]{Vogt2018}%
  \BibitemOpen
  \bibfield  {author} {\bibinfo {author} {\bibfnamefont {P.}~\bibnamefont
  {Vogt}}, \bibinfo {author} {\bibfnamefont {A.}~\bibnamefont {Mauze}},
  \bibinfo {author} {\bibfnamefont {F.}~\bibnamefont {Wu}}, \bibinfo {author}
  {\bibfnamefont {B.}~\bibnamefont {Bonef}},\ and\ \bibinfo {author}
  {\bibfnamefont {J.~S.}\ \bibnamefont {Speck}},\ }\bibfield  {title} {\bibinfo
  {title} {{Metal-oxide-catalyzed epitaxy (MOCATAXY): the example of O
  plasma-assisted molecular beam epitaxy of
  $\beta$-$(Al_xGa_{1-x})_2O_3/\beta-Ga_2O_3$ heterostructures}},\ }\href@noop
  {} {\bibfield  {journal} {\bibinfo  {journal} {Applied Physics Express}\
  }\textbf {\bibinfo {volume} {11}},\ \bibinfo {pages} {115503} (\bibinfo
  {year} {2018}{\natexlab{b}})}\BibitemShut {NoStop}%
\bibitem [{\citenamefont {Copel}\ \emph {et~al.}(1989)\citenamefont {Copel},
  \citenamefont {Reuter}, \citenamefont {Kaxiras},\ and\ \citenamefont
  {Tromp}}]{Copel1989}%
  \BibitemOpen
  \bibfield  {author} {\bibinfo {author} {\bibfnamefont {M.}~\bibnamefont
  {Copel}}, \bibinfo {author} {\bibfnamefont {M.}~\bibnamefont {Reuter}},
  \bibinfo {author} {\bibfnamefont {E.}~\bibnamefont {Kaxiras}},\ and\ \bibinfo
  {author} {\bibfnamefont {M.}~\bibnamefont {Tromp}},\ }\bibfield  {title}
  {\bibinfo {title} {{Surfactants in Epitaxial Growth}},\ }\href
  {https://journals-aps-org.proxy.lib.umich.edu/prl/pdf/10.1103/PhysRevLett.63.632}
  {\bibfield  {journal} {\bibinfo  {journal} {Physical Review Letters}\
  }\textbf {\bibinfo {volume} {63}},\ \bibinfo {pages} {632} (\bibinfo {year}
  {1989})}\BibitemShut {NoStop}%
\bibitem [{\citenamefont {Neugebauer}(2003)}]{Neugebauer2003b}%
  \BibitemOpen
  \bibfield  {author} {\bibinfo {author} {\bibfnamefont {J.}~\bibnamefont
  {Neugebauer}},\ }\bibfield  {title} {\bibinfo {title} {{Surfactants and
  antisurfactants on group-III-nitride surfaces}},\ }\href
  {https://doi.org/10.1002/pssc.200303132} {\bibfield  {journal} {\bibinfo
  {journal} {Physica Status Solidi C: Conferences}\ }\textbf {\bibinfo {volume}
  {0}},\ \bibinfo {pages} {1651} (\bibinfo {year} {2003})}\BibitemShut
  {NoStop}%
\bibitem [{\citenamefont {Lewis}\ \emph {et~al.}(2017)\citenamefont {Lewis},
  \citenamefont {Corfdir}, \citenamefont {Li}, \citenamefont {Herranz},
  \citenamefont {Pfuller}, \citenamefont {Brandt},\ and\ \citenamefont
  {Geelhaar}}]{Lewis2017}%
  \BibitemOpen
  \bibfield  {author} {\bibinfo {author} {\bibfnamefont {R.~B.}\ \bibnamefont
  {Lewis}}, \bibinfo {author} {\bibfnamefont {P.}~\bibnamefont {Corfdir}},
  \bibinfo {author} {\bibfnamefont {H.}~\bibnamefont {Li}}, \bibinfo {author}
  {\bibfnamefont {J.}~\bibnamefont {Herranz}}, \bibinfo {author} {\bibfnamefont
  {C.}~\bibnamefont {Pfuller}}, \bibinfo {author} {\bibfnamefont
  {O.}~\bibnamefont {Brandt}},\ and\ \bibinfo {author} {\bibfnamefont
  {L.}~\bibnamefont {Geelhaar}},\ }\bibfield  {title} {\bibinfo {title}
  {{Quantum Dot Self-Assembly Driven by a Surfactant-Induced Morphological
  Instability}},\ }\href {https://doi.org/10.1103/PhysRevLett.119.086101}
  {\bibfield  {journal} {\bibinfo  {journal} {Physical Review Letters}\
  }\textbf {\bibinfo {volume} {119}},\ \bibinfo {pages} {1} (\bibinfo {year}
  {2017})}\BibitemShut {NoStop}%
\bibitem [{\citenamefont {Madsen}\ and\ \citenamefont {Susi}(2021)}]{abtem}%
  \BibitemOpen
  \bibfield  {author} {\bibinfo {author} {\bibfnamefont {J.}~\bibnamefont
  {Madsen}}\ and\ \bibinfo {author} {\bibfnamefont {T.}~\bibnamefont {Susi}},\
  }\bibfield  {title} {\bibinfo {title} {{The abtem code: transmission electron
  microscopy from first principles.}},\ }\href
  {https://doi.org/10.12688/openreseurope.13015.1} {\bibfield  {journal}
  {\bibinfo  {journal} {Open Research Europe}\ }\textbf {\bibinfo {volume}
  {1}},\ \bibinfo {pages} {13015} (\bibinfo {year} {2021})}\BibitemShut
  {NoStop}%
\end{thebibliography}%
\end{document}